\documentclass[aps,prb,10pt,twocolumn,superscriptaddress,floatfix,longbibliography]{revtex4-2}
\usepackage{amsmath,amssymb,bm,graphicx,booktabs}
\usepackage[colorlinks=true,linkcolor=blue,citecolor=blue,urlcolor=blue]{hyperref}
\graphicspath{{./}}
\newcommand{\LL}{\mathcal L}
\newcommand{\JJ}{\boldsymbol{\mathcal J}}
\newcommand{\Tr}{\operatorname{Tr}}
\newcommand{\spec}{\operatorname{spec}}
\newcommand{\HS}{\mathrm{HS}}
\newcommand{\Id}{\mathbb I}
\newcommand{\youngtwotwo}{\vcenter{\hbox{\setlength{\unitlength}{8pt}%
\begin{picture}(2,2)\linethickness{0.4pt}%
\multiput(0,0)(0,1){3}{\line(1,0){2}}%
\multiput(0,0)(1,0){3}{\line(0,1){2}}%
\end{picture}}}}
\newcommand{\youngtwooneone}{\vcenter{\hbox{\setlength{\unitlength}{8pt}%
\begin{picture}(2,3)\linethickness{0.4pt}%
\multiput(0,0)(0,1){2}{\line(1,0){1}}%
\multiput(0,2)(0,1){2}{\line(1,0){2}}%
\multiput(0,0)(1,0){2}{\line(0,1){3}}%
\put(2,2){\line(0,1){1}}%
\end{picture}}}}
\hypersetup{pdftitle={A Lieb–Mattis theorem for Liouvillian decay spectra of isotropic spin jumps},pdfauthor={Daoguang Li, Kaixiang Lu, Mingdi Xu, Xiang-Ping Jiang, Lei Pan}}

\begin{document}
	\lefthyphenmin=3
	\righthyphenmin=3
	\title{A Lieb–Mattis theorem for Liouvillian decay spectra of isotropic spin jumps}
	\author{Daoguang Li}
	\affiliation{School of Physics, Nankai University, Tianjin 300071, China}
	\author{Kaixiang Lu}
	\affiliation{School of Physics, Nankai University, Tianjin 300071, China}
	\author{Mingdi Xu}
	\affiliation{School of Physics, Nankai University, Tianjin 300071, China}
	\author{Xiang-Ping Jiang}
\email[Corresponding author: ]{2015iopjxp@gmail.com}
\affiliation{School of Physics, Hangzhou Normal University, Hangzhou, Zhejiang 311121, China}
\author{Lei Pan}
\email[Corresponding author: ]{panlei@nankai.edu.cn}
\affiliation{School of Physics, Nankai University, Tianjin 300071, China}
\date{November 23, 2026}

\begin{abstract}
We establish a Lieb–Mattis ordering theorem for Liouvillian decay rates generated by isotropic triplets of Hermitian jump operators that are real linear combinations of local spins. Their Gram matrix defines an exchange magnet in operator space, where conserved local tensor ranks fix the spin lengths and a connected Marshall sign structure orders the sector edges above the sublattice-imbalance threshold. Every admissible graph admits a completely positive bond-jump realization, with spatially correlated magnetic noise providing a physical implementation. Separable two-spin preparations read out the hierarchy through local correlators, while spin-1/2 chains support Haldane-structured decay multiplets whose boundary order can reverse under frustration. Coherent-mixing bounds and collective depolarization quantify the stability and control of this ordering. Higher-spin, SU(4), and Hubbard examples connect these results to multipolar relaxation, Young-diagram sectors, and spin-resolved lifetimes. More broadly, by mapping microscopic noise correlations onto an operator-space exchange magnet, this work establishes a symmetry-based design principle for engineering and controlling long-lived decay channels in open many-body quantum systems.
\end{abstract}
\maketitle

\section{Introduction}
\label{sec:intro}

The relation between symmetry and spectral order is a central question in quantum magnetism, where rotational symmetry groups Hamiltonian eigenstates into total-spin multiplets whose relative energies reflect the interactions. The Lieb--Mattis (LM) theorem makes this connection precise: for a connected antiferromagnet with the appropriate bipartite sign structure, the lowest energy in each exact-spin sector increases with total spin above the ferrimagnetic threshold~\cite{Marshall1955,LiebMattis1962JMP}. By combining exchange signs, positivity, and angular-momentum addition, it determines the ground-state spin and a hierarchy of sector minima without solving the many-body spectrum. Related results for one-dimensional fermions and ferromagnetic chains~\cite{LiebMattis1962PR,NachtergaeleStarr2005}, together with Young-diagram ordering in SU($N$) systems~\cite{Hakobyan2004,Pan2017}, show how the relevant representation labels change with the symmetry and interactions. This perspective motivates a corresponding question for open systems: can the jump operators that describe dissipation determine which symmetry sectors relax most slowly? What is still missing is a sector-resolved ordering theorem that links the sign structure of microscopic jump operators to the lifetime hierarchy of Liouvillian decay modes, in the same way that the Lieb--Mattis theorem links exchange signs to energy ordering in closed magnets.

The Lindblad equation provides the natural setting for this question because its Liouvillian superoperator acts on density operators, with complex eigenvalues that specify the decay rates and oscillation frequencies of relaxation modes~\cite{Gorini1976,Lindblad1976,BreuerPetruccione}. These modes govern dissipative transitions, persistent motion, and boundary-sensitive relaxation~\cite{Minganti2018,BucaTindallJaksch2019,Mao2024}. In a rotationally covariant system, operator space decomposes into irreducible operator-spin sectors~\cite{BucaProsen2012,AlbertJiang2014}, whose smallest decay rates identify the longest-lived spectral channels. Comparing these edges gives an open-system counterpart of the LM problem, with angular momentum now carried by operators. Integrable constructions, operator-space fragmentation, and symmetry constraints have revealed substantial structure in Liouvillian spectra~\cite{Medvedyeva2016,EsslerPiroli2020,Nakagawa2021,Kawabata2024}, while superoperator bounds constrain their relaxation gaps~\cite{BarthelZhang2022}. An ordering theorem would further connect the lifetimes of different representations to the microscopic jump operators and explain how changing dissipation redistributes those lifetimes.

Here we establish such a theorem for isotropic triplets of Hermitian jumps built from real linear combinations of local spin components. The same spatial coefficients enter all three spin directions, and their Gram matrix generates an exchange magnet in operator space. Conserved local tensor ranks fix its adjoint spin lengths, while a connected Marshall sign structure orders the slowest multiplets within each rank pattern above the sublattice-imbalance threshold. This formulation identifies the dissipative class directly; spatially correlated magnetic noise gives an implementation of the jumps and a physical interpretation of their intersite exchange. We express the ordering condition as a graph test and construct a completely positive bond-jump realization for every admissible graph.

\begin{figure}[t]
  \centering
  \includegraphics[width=\columnwidth]{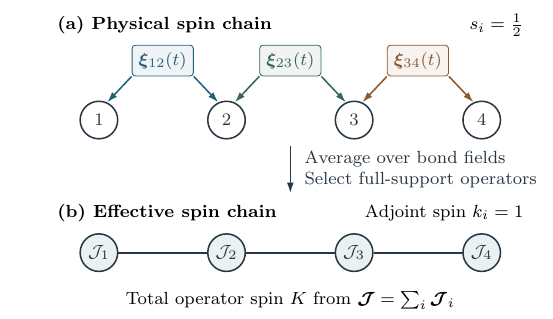}
  \caption{Four-site bond-noise model and its operator-space representation.
  (a) Independent isotropic bond fields $\boldsymbol\xi_{i,i+1}(t)$ couple with equal strength to their two endpoint spins, which have physical spin $s_i=1/2$.
  (b) Averaging the fields and selecting Pauli strings with a nonidentity factor at every site gives adjoint spins $k_i=1$, coupled by the antiferromagnetic exchange in Eq.~\eqref{eq:chain}. Their total operator spin $K$ labels the decay multiplets.}
  \label{fig:model}
\end{figure}

This perspective shifts dissipation from a merely decohering influence to a controllable source of exchange in operator space, where spatial noise correlations determine which collective modes survive longest. The physical consequences range from a local two-spin readout to spatially structured decay modes. Separable preparations expose the scalar, vector, and quadrupolar lifetimes through local correlators, while spin-$1/2$ chains embed a spin-$1$ antiferromagnet in the space of full-support operators, as illustrated in Fig.~\ref{fig:model}. Its decaying multiplets inherit Haldane string correlations and fractional boundary spins~\cite{Haldane1983,AKLT1987,WhiteHuse1993,Pollmann2010}, with frustration converting the known boundary crossings of the spin-$1$ chain~\cite{ChepigaMila2017} into reversals of decay order. We then establish coherent-mixing bounds and collective control of the global hierarchy, before connecting higher-spin multipolar channels to SU(4) exchange and Hubbard loss. The progression from physical spin to operator spin and Young-diagram sectors provides a common language for symmetry-resolved relaxation across these realizations. This work provides three results. First, we establish a Liouvillian Lieb–Mattis ordering theorem for isotropic Hermitian linear spin jumps, with the Gram matrix acting as an exchange magnet in operator space. Second, we show that every admissible exchange graph admits a completely positive bond-jump realization, including a spatially correlated magnetic-noise implementation. Third, we identify observable consequences in two-spin correlators, Haldane-structured decay multiplets of spin-1/2 chains, and frustration-induced reversals of the boundary decay order.

\section{Lieb--Mattis ordering in closed systems}
\label{sec:closed}

We first consider an SU(2)-invariant spin system, for which $[H,S_{\rm tot}^a]=0$ and total physical spin $S$ labels the energy multiplets. The LM theorem compares the lowest energy $E(S)$ among all multiplets of exact spin $S$; a fixed-magnetization block also contains higher-spin multiplets and therefore has minimum $\min_{S\geq|M|}E(S)$. Let $\mathbf T_i$ be a local spin operator with quantum number $t_i\in\{1/2,1,3/2,\ldots\}$, so $\mathbf T_i^2=t_i(t_i+1)\Id$ in units $\hbar=1$, and take the exchange Hamiltonian $H_{\rm LM}=\sum_{i<j}J_{ij}\mathbf T_i\cdot\mathbf T_j$. Suppose its nonzero interaction graph is connected and admits a partition $A\cup B$ with antiferromagnetic couplings across the partition and ferromagnetic couplings within either part. The ground multiplet then has the sublattice-imbalance spin $S_0=|\sum_{i\in A}t_i-\sum_{i\in B}t_i|$, above which the exact-spin minima obey~\cite{LiebMattis1962JMP,NachtergaeleStarr2005}
\begin{equation}
	E(S+1)>E(S),\qquad S\geq S_0,
	\label{eq:lm}
\end{equation}
whenever both sectors occur. For spin $1/2$, $S_0=|N_A-N_B|/2$, while arbitrary local spins enter through the weighted sublattice sums. A $\pi$ rotation of sublattice $A$ makes the off-diagonal exchange amplitudes nonpositive, so connectivity and angular-momentum addition fix the lowest vector's spin and establish the strict comparison~\cite{Marshall1955,LiebMattis1962JMP}. Total spin thus remains the ordering label for higher-spin SU(2) magnets satisfying these exchange conditions.

For SU($N$), the ordering instead concerns full irreducible representations. Consider an open chain of $L$ fundamental SU($N$) sites with $H=\sum_{i=1}^{L-1}J_iP_{i,i+1}$, where $P_{i,i+1}$ exchanges neighboring internal states and every $J_i>0$. Its sectors are labeled by Young diagrams $\lambda=(\lambda_1,\lambda_2,\ldots)$ with $L$ boxes and at most $N$ rows; rows encode symmetrization and columns antisymmetrization. Define dominance by $\lambda\succeq\mu$ when the sum of the first $r$ row lengths of $\lambda$ is at least that of $\mu$ for every $r$. Moving boxes downward produces a dominated diagram, and the generalized LM theorem orders the sector minima as~\cite{Hakobyan2004}
\begin{equation}
 E(\lambda)>E(\mu),\qquad \lambda\succeq\mu,\quad\lambda\ne\mu.
 \label{eq:sunorder}
\end{equation}
For example, moving one box from the second row to a third row gives, for $N\geq3$,
\[
 E\!\left(\youngtwotwo\right)
 >E\!\left(\youngtwooneone\right),
 \qquad (2,2)\succeq(2,1,1).
\]
For fundamental SU(2) sites only two rows occur, with $S=(\lambda_1-\lambda_2)/2$; additional rows at larger $N$ require the full diagram. Moreover, dominance is a partial order: $(4,1,1)$ and $(3,3)$ are incomparable because their first-row and first-two-row sums favor opposite diagrams. Specific integrable Fermi gases admit further comparisons through their Bethe spectra~\cite{Pan2017}, illustrating the additional dynamical information needed beyond representation dominance.

An enlarged-symmetry multiplet can contain several physical-spin values upon restriction to SU(2). For instance, when the SU(3) fundamental is implemented by a physical spin one, the $(2,2)$ multiplet contains $S=0,2$, whereas $(2,1,1)$ contains $S=1$. The preceding comparison therefore orders representation minima that cannot be identified by one total-spin label. The same distinction will arise in operator space: SU(2) covariance supplies an operator-spin label, while an SU($N$)-covariant Liouvillian is resolved by its full representation labels, as developed in Sec.~\ref{sec:young}.

\section{Master equation and operator-spin sectors}
\label{sec:ordering}

\subsection{Master equation and Liouvillian spectrum}

For a finite-dimensional system with Hamiltonian $H=H^\dagger$ and jump operators $L_\mu$, the Lindblad equation
\begin{equation}
	\frac{d\rho}{dt}=\LL(\rho)=-i[H,\rho]
	+\sum_\mu\left(L_\mu\rho L_\mu^\dagger
	-\frac12\{L_\mu^\dagger L_\mu,\rho\}\right)
	\label{eq:lindblad}
\end{equation}
defines a superoperator on $\mathcal B(\mathcal H)$ that generates completely positive, trace-preserving evolution~\cite{Gorini1976,Lindblad1976}. A right eigenoperator satisfies $\LL(R_\nu)=z_\nu R_\nu$, with $z_\nu=-\Gamma_\nu+i\omega_\nu$ and $\Gamma_\nu\geq0$, so its contribution to the initial state evolves as $e^{z_\nu t}R_\nu$. The real rate sets its lifetime and the imaginary part its oscillation frequency; stationary density operators belong to $z=0$, while Jordan blocks can add polynomial factors. We compare the smallest real decay rate in each invariant symmetry sector, which defines its decay edge and identifies its longest-lived spectral channel.

\subsection{From physical spin to operator spin}

For physical spins $\mathbf S_i$ with quantum numbers $s_i$, a common rotation acts on operators as $X\mapsto U_gXU_g^\dagger$. A covariant generator commutes with this rotation action, so its eigenoperators form spin multiplets. The infinitesimal local generators are the commutators $\mathcal J_i^aX=[S_i^a,X]$, whose sum $\JJ=\sum_i\JJ_i$ defines total operator spin by $\JJ^2X=K(K+1)X$. They obey the angular-momentum algebra and are Hermitian in the Hilbert--Schmidt inner product $\Tr(X^\dagger Y)$. For physical spin $1/2$, the components $S_i^+,S_i^z,S_i^-$ have magnetic weights $+1,0,-1$ and form an operator spin one, while the identity forms spin zero. For arbitrary physical spin, the local irreducible operator components have tensor ranks $k_i=0,\ldots,2s_i$, with
\begin{equation}
 \mathcal B(V_{s_i})=\bigoplus_{k_i=0}^{2s_i}V_{k_i},
 \qquad \mathcal M_{\bm k}=\bigotimes_iV_{k_i}.
 \label{eq:localranks}
\end{equation}
Thus the four nonidentity factors in Fig.~\ref{fig:model}(b) each have $k_i=1$ and combine into total operator spins $K=0,1,2,3,4$.

Within each operator-spin sector, any further conserved labels $r$ refine the decay edge to
\begin{equation}
	\Gamma_r(K)=-\max\{\operatorname{Re}z:z\in\spec(\LL|_{r,K})\}.
	\label{eq:general_edge}
\end{equation}
Covariance fixes the degeneracy among magnetic components, while $\Gamma_r(K)$ selects the longest-lived multiplet at fixed $r,K$ and $\Gamma(K)=\min_r\Gamma_r(K)$ gives the full-space edge. The linear spin jumps introduced next conserve each local rank, making $\bm k$ the additional label and turning its adjoint spins into the degrees of freedom of an exchange problem.

\section{Lieb--Mattis ordering for linear spin jumps}
\label{sec:theorem}

\subsection{Jump-operator class and exchange mapping}

The microscopic theorem concerns a purely dissipative Lindbladian, $H=0$, whose jumps occur in isotropic triplets,
\begin{equation}
 L_{\mu a}=\sum_i g_{\mu i}S_i^a,
 \qquad g_{\mu i}\in\mathbb R,\quad a=x,y,z.
 \label{eq:linearjumps}
\end{equation}
Here $\mu$ labels a dissipative channel, and the same spatial coefficients $g_{\mu i}$ enter its three spin components, making every jump Hermitian and the generator SU(2)-covariant. The coefficients include the square root of the jump rate. Their real Gram matrix $C_{ij}=\sum_\mu g_{\mu i}g_{\mu j}$ is positive semidefinite, and the Lindblad dissipator becomes $\LL_C(X)=-\tfrac12\sum_{ij,a}C_{ij}[S_i^a,[S_j^a,X]]$. Hence the decay operator has the exact exchange form
\begin{equation}
	-\LL_C=\frac12\sum_{ij}C_{ij}\JJ_i\cdot\JJ_j,
	\label{eq:map}
\end{equation}
which is positive semidefinite and Hilbert--Schmidt self-adjoint. Each local Casimir commutes with this operator, so the rank pattern $\bm k$ specifies an invariant module and its slowest exact-$K$ rate is $\Gamma_{\bm k}(K)$. The off-diagonal Gram entries supply the intersite exchange: with only one-site jumps, $C$ is diagonal and all total-spin sectors within a fixed-rank module have the same rate, whereas spatially extended jumps can split and order those multiplets.

Spatially correlated magnetic noise provides a realization of this class. Independent isotropic white fields coupled to the Hermitian channel components in Eq.~\eqref{eq:linearjumps} produce the same generator upon averaging; the induced site-field covariance is precisely $C$ (Appendix~\ref{app:algebra}). Thus the stochastic description gives a preparation mechanism for the dissipator defined above. For example, the bond jumps $L_{i,a}=\sqrt\gamma(S_i^a+S_{i+1}^a)$ can be generated by the common field at the two endpoints of each bond in Fig.~\ref{fig:model}(a). Each bond contributes $\gamma(\JJ_i+\JJ_{i+1})^2/2$ to the decay operator, giving $2\gamma\Id+\gamma\JJ_i\cdot\JJ_{i+1}$ when both local ranks are one. The four-site full-support module is therefore the spin-$1$ antiferromagnet in Fig.~\ref{fig:model}(b), with a common shift $6\gamma\Id$.

\subsection{Microscopic ordering theorem}

Only sites with $k_i>0$ are active in a module. Suppose there is a number $\chi\geq0$ such that the residual exchanges $J_{ij}=C_{ij}-\chi$ for $i\ne j$ form a connected graph on these sites and admit a partition $A\cup B$ with $J_{ij}\geq0$ across the partition and $J_{ij}\leq0$ within either part. Their sublattice-imbalance spin is $K_0=|\sum_{i\in A}k_i-\sum_{i\in B}k_i|$.

\paragraph*{Theorem (Liouvillian LM ordering).}
For $H=0$ and the Hermitian linear spin jumps in Eq.~\eqref{eq:linearjumps}, assume that their Gram matrix satisfies the connected exchange-sign condition above. Then the slowest rates in neighboring exact-$K$ sectors of the same local-rank module obey
\begin{equation}
	\Gamma_{\bm k}(K+1)-\Gamma_{\bm k}(K)
	>\chi(K+1)\geq0,\qquad K\geq K_0.
	\label{eq:theorem}
\end{equation}
Whenever both sectors occur, this inequality gives a strict hierarchy above $K_0$; for a balanced module, $K_0=0$, it begins with the singlet as the slowest multiplet.

\paragraph*{Proof.}
Insert the uniform shift $\chi$ into Eq.~\eqref{eq:map} to obtain
\begin{align}
	-\LL_C|_{\bm k}
	&=c_{\bm k}\Id+\frac{\chi}{2}\JJ^2+H_{\rm LM},
	\label{eq:decompose}\\
	c_{\bm k}&=\frac12\sum_i(C_{ii}-\chi)k_i(k_i+1),\notag
\end{align}
where $H_{\rm LM}=\sum_{i<j}J_{ij}\JJ_i\cdot\JJ_j$ has the Hamiltonian LM sign structure. Its lowest exact-$K$ energy therefore gives $\Gamma_{\bm k}(K)=c_{\bm k}+\chi K(K+1)/2+E_{\rm LM}(K)$, so subtracting neighboring sectors and applying Eq.~\eqref{eq:lm} proves Eq.~\eqref{eq:theorem}. Here $\chi$ is an algebraic shift of the Gram entries; adding physical collective jumps of strength $\delta\chi$ increases the adjacent-$K$ spacing by $\delta\chi(K+1)$. At $\chi=0$ the lowest multiplet of the entire module has spin $K_0$, while collective jumps can rearrange sectors below $K_0$ without changing the ordered sequence above it.\hfill$\square$

The condition has a direct graph interpretation: assign opposite sublattice signs across positive bonds and equal signs across negative bonds, propagating the assignment along a spanning tree and checking its consistency on the remaining edges~\cite{Harary1953}. Appendix~\ref{app:algebra} gives the equivalent cycle test. Conversely, every admissible exchange graph is realized by the bond jumps
\begin{equation}
	L_{ij,a}=\sqrt{|J_{ij}|}\,[S_i^a+\operatorname{sgn}(J_{ij})S_j^a],
	\label{eq:signedjump}
\end{equation}
together with collective jumps $\sqrt\chi\sum_iS_i^a$ and local jumps $\sqrt{\kappa_i}S_i^a$, with $\kappa_i\geq0$. Their Gram matrix has off-diagonal entries $C_{ij}=J_{ij}+\chi$ and is positive semidefinite by construction, while the local jumps shift all rates in a fixed-rank module equally. Correlated or anticorrelated endpoint fields implement the two bond signs, with the correlated case illustrated in Fig.~\ref{fig:model}.

\section{Two-spin and chain realizations}
\label{sec:models}

\begin{figure}[t]
	\centering
	\includegraphics[width=\columnwidth]{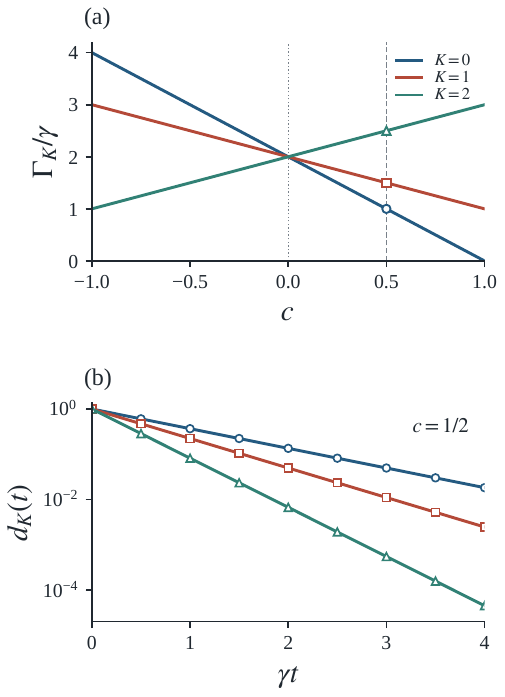}
	\caption{\label{fig:dimer}Spectral ordering and relaxation of two physical spin-$1/2$ sites. (a) Bilinear rates $\Gamma_K=\Gamma_{(1,1)}(K)$ for $C_{11}=C_{22}=\gamma$ and $C_{12}=\gamma c$; markers select $c=1/2$ for (b). (b) Normalized steady-state distances $d_K(t)$ for the separable preparations in Eq.~\eqref{eq:dimerreadout}, with equal initial distance $\epsilon_{\rm p}=1/4$. Lines show exact exponentials and symbols full Lindblad evolution. Colors identify the scalar ($K=0$), vector ($K=1$), and quadrupolar ($K=2$) channels in both panels.}
\end{figure}

\subsection{Two-spin hierarchy and relaxation}

The simplest realization uses two physical spin-$1/2$ sites with jumps $L_{\pm,a}=\sqrt{\gamma(1\pm c)/2}(S_1^a\pm S_2^a)$, where $-1\leq c\leq1$. Their Gram matrix has $C_{11}=C_{22}=\gamma$ and $C_{12}=\gamma c$, interpolating between correlated and anticorrelated fluctuations. The bilinear module $\bm k=(1,1)$ contains scalar, vector, and quadrupolar components with $K=0,1,2$ and rates
\begin{equation}
	\frac{\Gamma_{(1,1)}(K)}{\gamma}
	=2+\frac c2[K(K+1)-4],
	\label{eq:dimer}
\end{equation}
which increase with $K$ for $c>0$, coincide under independent noise ($c=0$), and reverse for $c<0$ [Fig.~\ref{fig:dimer}(a)]. Including the stationary identity and one-site modes of rate $\gamma$ gives the full-space edges $\Gamma(0)=0$, $\Gamma(1)=\gamma$, and $\Gamma(2)=\gamma(2+c)$, so the bilinear hierarchy resolves a correlation-dependent ordering within this nondecreasing global envelope.

To connect this spectrum to state relaxation, take the Hermitian scalar, vector, and quadrupolar operators $B_K$ of unit Hilbert--Schmidt norm defined in Appendix~\ref{app:dimerreadout}. The separable preparations $\rho_K(0)=\Id/4+\epsilon_{\rm p}B_K$, with $0<\epsilon_{\rm p}\leq1/(2\sqrt3)$, have equal Hilbert--Schmidt distance $\epsilon_{\rm p}$ from the common steady state $\rho_{\rm ss}=\Id/4$ for $|c|<1$. Each deviation $\rho_K(0)-\rho_{\rm ss}$ belongs to the fixed bilinear sector $(\bm k,K)=((1,1),K)$, which is preserved throughout the evolution. The corresponding local readouts obey
\begin{align}
	s_K(t)&=\Tr[B_Ke^{\LL_Ct}\rho_K(0)]
	=\epsilon_{\rm p}e^{-\Gamma_{(1,1)}(K)t},\notag\\
	\frac{s_1(t)}{s_0(t)}&=e^{-\gamma ct},\qquad
	\frac{s_2(t)}{s_1(t)}=e^{-2\gamma ct}.
	\label{eq:dimerreadout}
\end{align}
Figure~\ref{fig:dimer}(b) displays the normalized distances $d_K(t)=\|\rho_K(t)-\rho_{\rm ss}\|_{\HS}/\epsilon_{\rm p}=s_K(t)/s_K(0)$, with direct Lindblad evolution confirming the analytic curves. At positive $c$, the smaller scalar decay rate leaves the largest deviation at every positive time, followed by the vector and quadrupolar channels. Each selected sector contains one multiplet, whose rate governs the entire relaxation toward the common steady state. These different time scales can be read through two or three local Pauli correlators per preparation, while extending the construction to chains introduces multiple rates within a sector and spatially structured decay modes.

\begin{figure*}[t]
	\centering
	\includegraphics[width=\textwidth]{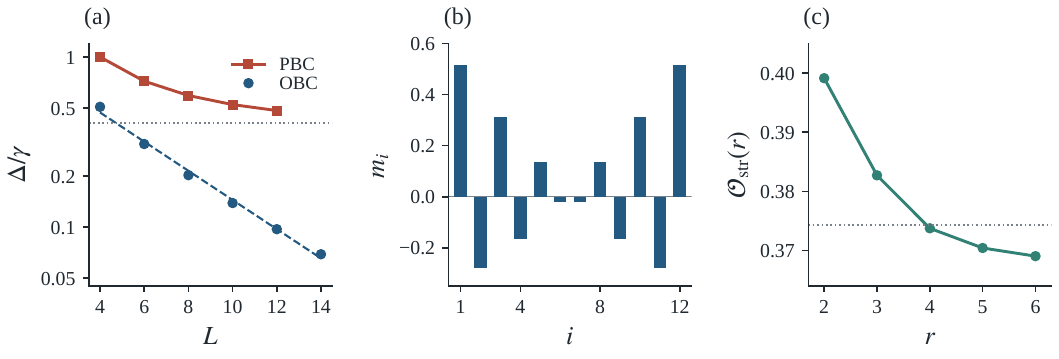}
	\caption{\label{fig:haldane}Haldane structure of the full-support decay eigenoperators for nearest-neighbor bond noise. (a) Singlet--triplet separation with open and periodic boundaries, together with an exponential fit of the open-chain data for $L=4$--$14$ and the thermodynamic spin-$1$ gap as a periodic-chain reference. (b) Adjoint-spin profile of the $L=12$ open-chain triplet with $M=1$, showing boundary localization and staggered tails, with $\sum_i m_i=1$. (c) String correlation of the $L=12$ periodic singlet compared with its thermodynamic benchmark~\cite{WhiteHuse1993}. Symbols denote finite-size calculations.}
\end{figure*}

\subsection{Bond jumps and Haldane decay multiplets}
\label{sec:haldane}

For an open or periodic chain with nearest-neighbor jumps $L_{i,a}=\sqrt\gamma(S_i^a+S_{i+1}^a)$, the full-support module $\bm1=(1,\ldots,1)$ has decay operator
\begin{equation}
	-\LL_C|_{\bm1}=2\gamma N_b\Id+\gamma H_1,
	\qquad H_1=\sum_{\langle ij\rangle}\JJ_i\cdot\JJ_j,
	\label{eq:chain}
\end{equation}
where $N_b=L-1$ for open boundaries and $N_b=L$ for periodic boundaries. The effective spin-$1$ exchange acts on the nonidentity local operators of the physical spin-$1/2$ sites; for even $L$, its balanced graph orders the sector edges strictly from the singlet to maximal operator spin. The common offset $2\gamma N_b$ sets the absolute decay scale, while rate differences equal $\gamma$ times the spin-$1$ energy differences, yielding a finite periodic singlet--triplet separation set by the Haldane gap~\cite{WhiteHuse1993}. With open boundaries, two effective spin-$1/2$ degrees of freedom form a nearly degenerate singlet and triplet, whose splitting decreases as $\gamma A e^{-L/\xi_{\rm ED}}$ over the sizes in Fig.~\ref{fig:haldane}(a).

The boundary profile and bulk string correlation reveal the spatial structure of these normalized eigenoperators (definitions in Appendix~\ref{app:numerics}). The triplet profile $m_i$ is concentrated at the open ends with staggered tails, while the periodic singlet has a nonzero string correlator $\mathcal O^z_{\rm str}(r)$ approaching the spin-$1$ benchmark~\cite{WhiteHuse1993} [Figs.~\ref{fig:haldane}(b) and \ref{fig:haldane}(c)]. These structures belong to decaying many-spin modes around the unique maximally mixed steady state $\rho_{\rm ss}=\Id/2^L$, whose uniqueness for $L\geq3$ follows because overlapping bond generators produce every local spin component (Appendix~\ref{app:algebra}). The integer-spin onsite SO(3) action and projective spin-$1/2$ boundaries of the edge eigenoperators~\cite{AKLT1987,Pollmann2010} connect this construction to open-system and doubled-space approaches~\cite{deGroot2022,Verissimo2023,Paszko2024,MaTurzillo2025}. Lower-support operators have their own decay scales, including rates $\gamma$ for an open-chain endpoint spin and $2\gamma$ for a periodic one-site spin; Sec.~\ref{sec:controls} describes collective control of the full-space edges across these support patterns.

\begin{figure*}[t]
	\centering
	\includegraphics[width=\textwidth]{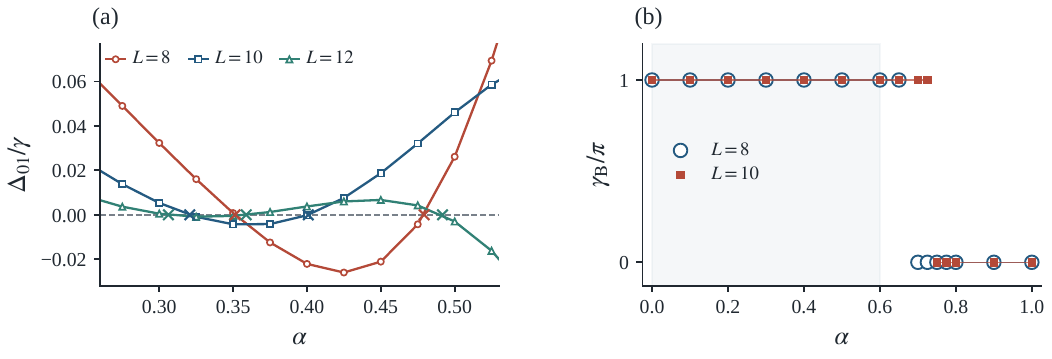}
	\caption{\label{fig:frustration}Boundary ordering and bond Berry phase in the completely positive frustrated chain. (a) Open-chain singlet--triplet margin for $L=8,10,12$, with crosses marking the refined zero crossings. (b) Bond Berry phase for periodic $L=8,10$, evaluated on a discretized twist loop in the $M=0$ block. The phase remains $\pi$ at the sampled points throughout the shaded ordering-scan interval and changes to zero in a size-dependent interval at larger frustration, near the known transition of the spin-$1$ chain~\cite{KolezhukSchollwoeck2002}. Lines join sampled points within each branch.}
\end{figure*}

\subsection{Boundary decay-rate crossings under frustration}
\label{sec:frustration}

The boundary hierarchy can be tuned by adding next-nearest-neighbor jumps $L^{(2)}_{i,a}=\sqrt\eta(S_i^a+S_{i+2}^a)$, for which $\alpha=\eta/\gamma\geq0$ gives
\begin{align}
	-\LL_C|_{\bm1}
	&=2(\gamma N_1+\eta N_2)\Id+\gamma(H_1+\alpha H_2),
	\label{eq:j1j2}\\
	H_2&=\sum_{\langle\langle ij\rangle\rangle}\JJ_i\cdot\JJ_j.\notag
\end{align}
Here $N_1,N_2$ count the respective bonds, with $N_2=L-2$ for open boundaries and $N_2=L$ for the periodic sizes studied here. Throughout this completely positive family, antiferromagnetic triangles frustrate the nearest-neighbor Marshall partition for $\alpha>0$, while finite-size spectral continuity preserves an initial interval of ordering near $\alpha=0$. At larger frustration, the open-chain singlet--triplet margin $\Delta_{01}=\Gamma_{\bm1}(1)-\Gamma_{\bm1}(0)$ changes sign repeatedly [Fig.~\ref{fig:frustration}(a)], reflecting the interaction $J_{\rm edge}\mathbf s_L\cdot\mathbf s_R$ between the effective boundary spins. Its exponentially small strength can change sign within the Haldane regime, as in spin-$1$ chains with incommensurate correlations~\cite{KolezhukRothSchollwoeck1997,ChepigaMila2017}.

The periodic chain supplies a complementary bulk diagnostic: twisting one nearest-neighbor bond and following the lowest nondegenerate $M=0$ state around the loop gives a bond Berry phase quantized by time reversal to $0$ or $\pi$~\cite{Hatsugai2006Berry} (Appendix~\ref{app:berry}). It remains $\pi$ at the sampled points in the open-chain crossing region and changes only at larger frustration [Fig.~\ref{fig:frustration}(b)]. Together with the finite periodic relative gap and nonzero string correlations across the same scan, this identifies the observed crossings with a reversal of the boundary coupling within the Haldane regime. The perturbation bound in Appendix~\ref{app:stability} quantifies the initial ordered interval beyond the Marshall condition, while the dynamical readout below relates the sector edges to relaxation signals.

\subsection{Symmetry-resolved relaxation}
\label{sec:dynamics}

To read out the sector edge in time, choose a probe $O_K$ of unit Hilbert--Schmidt norm at fixed $\bm k$ and exact $K$. Expanding it in the orthonormal decay eigenoperators gives
\begin{align}
	C_K(t)&=\langle O_K,e^{\LL_Ct}O_K\rangle_{\HS}
	=\sum_n w_{K,n}e^{-\Gamma_{\bm k,K,n}t},
	\label{eq:correlation}\\
	w_{K,n}&\geq0,\qquad \sum_nw_{K,n}=1.\notag
\end{align}
The logarithmic derivative defines the instantaneous rate, $\Gamma_K^{\rm eff}=-\partial_t\ln C_K$, which tends to $\Gamma_{\bm k}(K)$ whenever the edge has nonzero weight. It approaches monotonically from above because its derivative is minus the rate variance under the normalized time-dependent weights, so the LM theorem orders the limiting rates of the symmetry-resolved signals. Their approach to a single exponential is bounded by
\begin{equation}
	0\leq\frac{C_K(t)}{w_K e^{-\Gamma_{\bm k}(K)t}}-1
	\leq\frac{1-w_K}{w_K}e^{-\delta_Kt}.
	\label{eq:visibilitybound}
\end{equation}
Here $0<w_K<1$ is the total edge weight and $\delta_K>0$ the separation to the next supported rate, so the edge overlap and the spacing within the sector jointly determine the onset of single-rate relaxation; when $w_K=1$, the signal is a single exponential at all times.

\begin{figure*}[t]
	\centering
	\includegraphics[width=\textwidth]{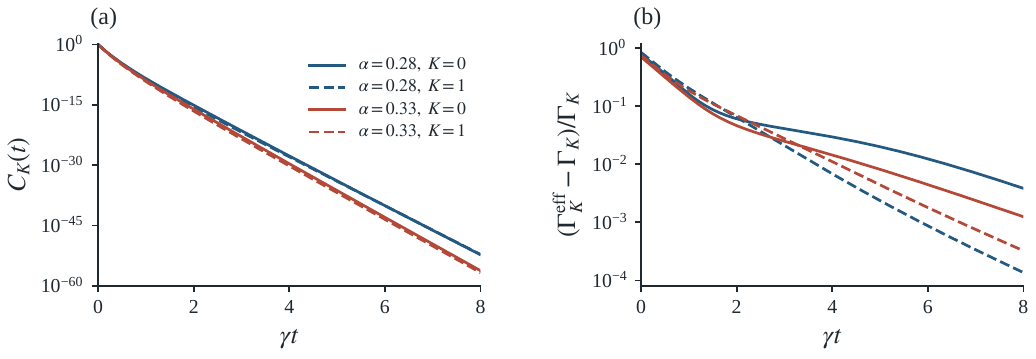}
	\caption{\label{fig:dynamics}Relaxation of a fixed projected Pauli-string probe for $L=12$. (a) Normalized correlations $C_K(t)$ at $\alpha=0.28$ and $0.33$, on opposite sides of the first boundary crossing. Solid and dashed lines denote $K=0$ and $K=1$. (b) Relative excess of the instantaneous rate above its sector edge, $[\Gamma_K^{\rm eff}(t)-\Gamma_K]/\Gamma_K$, with the same colors and line styles. All four rates approach their edges monotonically from above. The small boundary splitting and the extensive common decay scale are distinct; the latter sets the rapid decrease in (a). Here $\Gamma_K=\Gamma_{\bm1}(K)$.}
\end{figure*}

To compare lengths and frustrations with the same probe, we select the full-support module using a Pauli string with a nonidentity factor at every site and project its repeating-axis pattern onto the highest-weight component $M=K$. Appendix~\ref{app:probe} gives the projector and the corresponding product-state signal. Figure~\ref{fig:dynamics} follows the resulting instantaneous rates toward their sector edges on either side of the first boundary crossing: the extensive offset in Eq.~\eqref{eq:j1j2} causes the correlations to fall over many decades, while the much weaker end-spin interaction determines the singlet--triplet splitting. For this probe the asymptotic signal is small, as set by both its edge overlap and the preparation factor in Appendix~\ref{app:probe}. Increasing the edge weight at fixed $\delta_K$ brings single-rate relaxation to earlier times according to Eq.~\eqref{eq:visibilitybound}; resolving the boundary splitting additionally requires the residual shifts of both instantaneous rates to fall below that splitting.

\section{General criteria, stability, and global ordering}
\label{sec:controls}

\subsection{Hermitian representative and exchange criterion}
\label{sec:general}

The exchange mechanism of Sec.~\ref{sec:theorem} extends to generators with a symmetry-preserving Hermitian representative. Let an invariant module $r$ carry a tensor product of effective SU(2) spins $j_i>0$, and suppose a positive metric $G_r$ commutes with spin rotations and satisfies $G_r\LL_r=\LL_r^\dagger G_r$. The similar operator $\widetilde H_r=-G_r^{1/2}\LL_rG_r^{-1/2}$ is then Hermitian and shares the sector eigenvalues of $-\LL_r$. Such metrics occur in reversible and detailed-balance dynamics~\cite{Davies1974,ChenKastoryanoGilyen2023}, allowing their ordering to be related to an exchange structure through the following criterion.

\paragraph*{Proposition (Liouvillian LM criterion).}
Assume that, up to a unitary preserving the SU(2) action,
\begin{equation}
	\widetilde H_r=c_r+f_r(\JJ^2)+\sum_{i<j}J_{ij}\JJ_i\cdot\JJ_j,
	\label{eq:generalrep}
\end{equation}
where the exchange graph has the connected LM sign structure of Sec.~\ref{sec:closed}, and $f_r$ is nondecreasing on the allowed Casimir values. Then $\Gamma_r(K+1)>\Gamma_r(K)$ for every occurring $K\geq K_0$, with $K_0=|\sum_Aj_i-\sum_Bj_i|$.

Indeed, $\Gamma_r(K)=c_r+f_r[K(K+1)]+E_{\rm LM}(K)$, where $E_{\rm LM}(K)$ is the lowest exchange energy, so the strict comparison~\eqref{eq:lm} and the monotonicity of $f_r$ give the result, with the singlet slowest in a balanced module. The linear-jump theorem is its $G_r=\Id$ realization, through Eq.~\eqref{eq:decompose}.

\subsection{Stability under coherent mixing}

A coherent term that is anti-Hermitian in the metric of Sec.~\ref{sec:general}, preserves $r,K$, and commutes with the dissipator changes only the frequencies, since the two contributions can be diagonalized together. When they do not commute, their mixing can also shift decay rates. To control this shift, take a Hermitian SU(2)-covariant $D=-\LL_{\rm diss}$ and an SU(2)-invariant Hamiltonian $H$, with $A=[H,\,\cdot\,]$ and $-\LL_\varepsilon=D+i\varepsilon A$. In an exact-$K$ block, let $P_K$ project onto the complete lowest eigenspace of $D_K$, whose eigenvalue $d_K$ is separated from the remaining spectrum by $g_K>0$. Setting $Q_K=\Id-P_K$ and $b_K=|\varepsilon|\|Q_KA_KP_K\|$, the condition $b_K<g_K/2$ gives
\begin{equation}
	0\leq\Gamma_\varepsilon(K)-d_K\leq r_K
	=\frac{g_K-\sqrt{g_K^2-4b_K^2}}2
	\leq\frac{2b_K^2}{g_K}.
	\label{eq:coherentbound}
\end{equation}
An initial spacing $d_{K+1}-d_K>r_K$ therefore retains its sign, even with arbitrarily strong coherent dynamics inside either diagonal block, because the bound controls the coupling out of the slow eigenspace. Appendix~\ref{app:coherent} proves this statement by isolating a vertical spectral strip. Forming the projectors in the full exact-$K$ space allows $H$ to mix local ranks, while any further module invariant under both contributions admits the same argument.

\subsection{Collective control of global decay ordering}
\label{sec:collective}

Collective jumps control the global edges by shifting every mode of a given $K$ equally. For any finite-dimensional SU(2)-covariant Lindblad generator $\LL_0$, adding three jumps $\sqrt\chi S_{\rm tot}^a$, with $\chi\geq0$, gives $\LL_\chi=\LL_0-\chi\JJ^2/2$. This isotropic collective depolarization~\cite{RivasLuis2013,DenisMartin2022} acts as a scalar Casimir shift in each exact-$K$ space, including all multiplicity labels, so that
\begin{align}
	z_{K,n}(\chi)&=z_{K,n}(0)-\frac\chi2K(K+1),\notag\\
	\Gamma_\chi(K)&=\Gamma_0(K)+\frac\chi2K(K+1),
	\label{eq:collectiveshift}
\end{align}
while preserving the sector eigenoperators, Jordan chains, and oscillation frequencies. The scalar block $K=0$ remains unchanged and every nonscalar block acquires additional decay, with the identities applying equally to nonnormal and nonunital backgrounds.

\paragraph*{Corollary (global ordering by collective noise).}
For adjacent occurring operator spins, let
\begin{equation}
	\chi_* =\max_K
	\frac{[\Gamma_0(K)-\Gamma_0(K+1)]_+}{K+1},
	\qquad [x]_+=\max(x,0).
	\label{eq:collectivethreshold}
\end{equation}
Then $\chi>\chi_*$ gives $\Gamma_\chi(K+1)>\Gamma_\chi(K)$, because each adjacent spacing is shifted by exactly $\chi(K+1)$, independently of the multiplicity label. This comparison therefore survives minimization over invariant modules. For the integer operator spins of a tensor-product spin system, the sufficient threshold $\chi>\sum_\mu\|L_\mu\|_{\rm op}^2$ follows directly from any jump representation of $\LL_0$. Indeed, writing $R=2\sum_\mu\|L_\mu\|_{\rm op}^2$, the Hilbert--Schmidt expectation of the eigenvalue equation gives $0\leq\Gamma_n\leq R$, since the Hamiltonian commutator contributes no real part and each dissipator has induced norm at most $2\|L_\mu\|_{\rm op}^2$. Rotationally averaging a stationary state gives $\Gamma_0(0)=0$, making the $0\to1$ spacing at least $\chi$ and every $K\geq1$ spacing at least $-R+\chi(K+1)>0$. The threshold is uniform over all SU(2)-invariant Hamiltonians at fixed dissipators, including strong interactions that mix local ranks, and complements bounds on the overall gap from local noise in unital dynamics~\cite{ZhangBarthel2026}.

\subsection{Global edges from the jump coefficients}

For $L$ physical spin-$1/2$ sites with linear jumps~\eqref{eq:linearjumps}, the smallest and largest Gram-matrix eigenvalues $c_-$ and $c_+$ give bounds that include all local-rank modules:
\begin{equation}
	Kc_-\leq\Gamma(K)\leq Kc_+,\qquad K=0,\ldots,L.
	\label{eq:globalcovbounds}
\end{equation}
The lower bound follows because a module with $q$ active sites has every rate at least $qc_-$ and contains spin $K$ only if $q\geq K$. For the upper bound, the highest-weight string on any $K$ sites is an exact spin-$K$ eigenoperator with rate at most $Kc_+$ (Appendix~\ref{app:algebra}). Adjacent edges therefore obey $\Gamma(K+1)-\Gamma(K)\geq(K+1)c_--Kc_+$, so for $L\geq2$ and $c_->0$ the condition $c_+/c_-<L/(L-1)$ guarantees strict global ordering. Independent equal-strength jumps, for which $C=c\Id$, saturate both bounds with $\Gamma(K)=Kc$; a finite neighborhood of that limit inherits the hierarchy. The Gram spectrum thus provides a global criterion complementary to the exchange-sign comparison within fixed-rank modules.

\section{Higher spins, enlarged symmetry, and particle loss}
\label{sec:higher}

\subsection{Multipolar channels of higher-spin SU(2) systems}

For physical spin $s>1/2$, the ranks $k=0,1,\ldots,2s$ in Eq.~\eqref{eq:localranks} supply scalar, dipolar, quadrupolar, and higher multipolar channels. Each fixed pattern remains a tensor product of adjoint SU(2) spins, so the same linear-jump theorem applies with total $K$ as the eigenoperator label. In the two-site model of Sec.~\ref{sec:models}, angular-momentum addition gives adjacent spacings $\gamma c(K+1)$ for arbitrary local ranks, ordering all five multiplets of the spin-$1$ quadrupole--quadrupole module $(2,2)$ when $c>0$.

Higher spins also allow the tensor character of the jumps themselves to be varied. A different dissipative class is supplied by the rank-two jumps $L_m=\sqrt\kappa\,T_m^{(2)}$ on one physical spin-$1$ site, with $m=-2,\ldots,2$ and Hilbert--Schmidt-orthonormal tensors, which give the full-sector rates
\begin{equation}
	\Gamma(0)=0,\qquad \Gamma(1)=\frac{5\kappa}{2},\qquad
	\Gamma(2)=\frac{3\kappa}{2},
	\label{eq:quadrupolarrates}
\end{equation}
so the dipole decays faster than the quadrupole. These rates follow by completing $S^a/\sqrt2$ with the five rank-two tensors to a traceless orthonormal basis, whose full dissipator is $X\mapsto\Tr(X)\Id-3X$, and subtracting its rank-one contribution. Adding magnetic jumps $\sqrt\chi S^a$ then changes the spacing to $\Gamma_\chi(2)-\Gamma_\chi(1)=2\chi-\kappa$, restoring strict global ordering at $\chi>\kappa/2$ and providing an exact example of collective control over a reversed hierarchy.

\subsection{Young-diagram sectors and the SU(4) example}
\label{sec:young}

For an SU($N$)-covariant generator, operator space decomposes into full irreducible representations, with decay edge $\Gamma(\lambda)=-\max\operatorname{Re}\spec(\LL_\lambda)$ in each Young-diagram sector. This is the open-system counterpart of Sec.~\ref{sec:closed}: a symmetry-preserving Hermitian representative inherits the dominance comparison~\eqref{eq:sunorder} when its representations and exchange couplings realize the antiferromagnetic chain specified there. The relevant local representation is that of operators; for a physical fundamental SU($N$) site, it contains a singlet and an adjoint multiplet. Matching these representations and the exchange signs determines which Hamiltonian ordering applies.

A concrete enlarged symmetry arises from bilinear exchange jumps between physical spin-$1/2$ sites. Taking $h_i=\mathbf S_i\cdot\mathbf S_{i+1}$ and $L_i=\sqrt{2\kappa}\,h_i$ with $H=0$, the identity $h_i=P_{i,i+1}/2-\Id/4$ gives
\begin{equation}
	\LL_{\rm ex}=\frac\kappa2\sum_i(\Pi_{i,i+1}^{(4)}-\Id),
	\label{eq:su4}
\end{equation}
where $\Pi^{(4)}$ swaps the four-dimensional onsite operator spaces, which carry the fundamental representation of this emergent SU(4). Here $-\LL_{\rm ex}$ is a ferromagnetic permutation Hamiltonian~\cite{Sutherland1975}, favoring symmetric diagrams: for two sites the symmetric diagram $(2)$ has rate zero and the antisymmetric diagram $(1,1)$ has rate $\kappa$. The fully symmetric representation remains stationary on a connected chain, whose uniform periodic form also has the exact color-magnon branch $\Gamma(q)=\kappa(1-\cos q)$. Branching the SU(4) representations into physical SU(2)$_L\times$SU(2)$_R$ sectors and minimizing the resulting multiplicity spectra gives the physical-spin decay edges (Appendix~\ref{app:su4}). This example makes explicit how the jump operators select both the representation labels and the exchange sign governing relaxation.

\subsection{Spin-preserving loss: the Hubbard model}
\label{sec:hubbard}

When $H$ and every $L_\mu$ commute with physical total spin, rotations act independently on the two Liouville legs, giving labels $(S_L,S_R)$ and a common multiplicity-space eigenvalue throughout $|S_L-S_R|\leq K\leq S_L+S_R$~\cite{BucaProsen2012,AlbertJiang2014} (Appendix~\ref{app:strong}). This stronger symmetry resolves both the preceding exchange model and the Fermi--Hubbard chain with onsite singlet loss,
\begin{align}
	H_{\rm Hub}&=-t_{\rm h}\sum_{i,\sigma}(c_{i\sigma}^\dagger c_{i+1,\sigma}+\mathrm{H.c.})
	+U\sum_i n_{i\uparrow}n_{i\downarrow},\notag\\
	L_i&=\sqrt\kappa\,c_{i\downarrow}c_{i\uparrow},
	\label{eq:hubbard}
\end{align}
which is relevant to two-body-loss experiments in ultracold fermions~\cite{Sponselee2019}. The scalar jumps conserve physical total spin while removing particle pairs, producing the no-jump Hubbard Hamiltonian at complex interaction $U_{\rm c}=U-i\kappa/2$~\cite{Dalibard1992,LiebWu1968,Nakagawa2021}. Since pair recycling lowers particle number on both Liouville legs, the full generator is triangular and has diagonal-block eigenvalues
\begin{equation}
	z_{ab}=-i(E_a^{\rm eff}-E_b^{{\rm eff}*}),\qquad
	\Gamma_{ab}=\frac{\kappa_a+\kappa_b}{2},
	\label{eq:hubbardwidth}
\end{equation}
where $\kappa_a=-2\operatorname{Im}E_a^{\rm eff}$ is a one-leg loss width. The labels $(N_L,S_L;N_R,S_R)$ therefore organize spin-resolved lifetimes, with dark fully polarized sectors for $N\leq L$ and a quadratic decay dispersion on the lowest spin-wave branch. Comparing the minimized widths across physical-spin sectors defines the associated ordering problem, which can be approached through the complex-interaction Bethe equations and spin-wave expansion in Appendix~\ref{app:hubbard}.

\section{Conclusions}

We have established a Lieb--Mattis hierarchy of Liouvillian decay edges for isotropic Hermitian jumps that are real linear combinations of local spins, whenever their Gram matrix has the connected Marshall exchange-sign structure. Conserved local tensor ranks determine the adjoint spins, whose multiplets are ordered above the sublattice-imbalance threshold. Every admissible signed graph admits a completely positive bond-jump realization, with spatially correlated magnetic fields providing a physical implementation. The hierarchy admits a local readout through separable two-spin preparations, while chains support Haldane-structured decay modes around a unique maximally mixed steady state. Their fractional boundary spins acquire a frustration-dependent coupling whose sign reversals change the singlet--triplet decay order within the sampled $\pi$-Berry-phase regime; the spectral overlap of a probe then determines when this structure enters its relaxation signal.

The stability and control results extend this picture beyond a fixed Hermitian exchange problem: coherent mixing shifts an isolated decay edge by an amount bounded quadratically in its coupling to higher modes; collective depolarization imposes a global ordering above a Hamiltonian-independent sufficient threshold; and the Gram spectrum certifies a global hierarchy near independent spin-$1/2$ jumps. Higher-spin multipolar channels, SU($N$) exchange, and Hubbard loss expose the roles of the jumps' tensor character, enlarged-symmetry Young diagrams, and physical-spin conservation.  A natural next step is to develop microscopic ordering criteria for nonunital SU($N$) dynamics and to design probes that retain appreciable overlap on many-body edge modes, offering a route from these spectral relations to controlled relaxation in a broader class of open quantum systems. More broadly, the operator-space exchange picture turns noise correlations into a design principle for symmetry-resolved dissipation and long-lived collective modes in open many-body quantum systems.

\section*{Data availability}
The data and source code supporting the figures and finite-size calculations accompany the manuscript.

\section*{Acknowledgment}
This work was supported by the National Natural Science Foundation of China (Grant Nos. 12674355, 12505017 and 12304290), Beijing National Laboratory for Condensed Matter Physics (Grant No. 2025BNLCMPKF017), and the Fundamental Research Funds for the Central Universities (010-63263115).

\appendix
\section{Operator modules and the sign condition}
\label{app:algebra}

The jump class in Eq.~\eqref{eq:linearjumps} has a stochastic realization through $H_\xi(t)=\sum_{\mu,a}\zeta_\mu^a(t)L_{\mu a}$, where the real Gaussian fields have zero mean and $\overline{\zeta_\mu^a(t)\zeta_\nu^b(t')}=\delta_{\mu\nu}\delta_{ab}\delta(t-t')$. The effective field on site $i$ is $\xi_i^a=\sum_\mu g_{\mu i}\zeta_\mu^a$, whose covariance is $\overline{\xi_i^a(t)\xi_j^b(t')}=C_{ij}\delta_{ab}\delta(t-t')$. Averaging the unitary white-noise evolution gives $\LL_C(X)=-\tfrac12\sum_{ij,a}C_{ij}[S_i^a,[S_j^a,X]]$, recovering the dissipator defined by the jumps. Conversely, diagonalizing any real positive semidefinite $C=\sum_\mu c_\mu v_\mu v_\mu^{\mathsf T}$ supplies real jump coefficients $g_{\mu i}=\sqrt{c_\mu}(v_\mu)_i$, establishing the equivalence of the two descriptions.

Choose local irreducible tensors $T^{(k)}_m$ normalized by $\Tr[(T^{(k)}_m)^\dagger T^{(k')}_{m'}]=\delta_{kk'}\delta_{mm'}$, whose commutators
\begin{align}
	[S^z,T^{(k)}_m]&=mT^{(k)}_m,\notag\\
	[S^\pm,T^{(k)}_m]&=\sqrt{k(k+1)-m(m\pm1)}\,T^{(k)}_{m\pm1}
	\label{eq:tensors}
\end{align}
identify the adjoint action on their product basis as a tensor product of spins $k_i$. These products form an orthonormal basis of $\mathcal M_{\bm k}$, while cyclicity of the trace establishes self-adjointness,
\begin{equation}
	\Tr\{X^\dagger[S_i^a,Y]\}
	=\Tr\{[S_i^a,X]^\dagger Y\},
	\label{eq:selfadj}
\end{equation}
which, together with the sum-of-squares form $-\LL_C=\frac12\sum_{\mu,a}c_\mu(\sum_i(v_\mu)_i\mathcal J_i^a)^2$ obtained by diagonalizing $C$, proves nonnegativity of the decay operator.

Write the two sublattice signs as $\eta_i=\pm1$, so that each nonzero exchange satisfies $\operatorname{sgn}J_{ij}=-\eta_i\eta_j$. A consistent assignment exists exactly when every cycle $\mathcal C$ obeys
\begin{equation}
 \prod_{(ij)\in\mathcal C}\operatorname{sgn}J_{ij}=(-1)^{|\mathcal C|}.
 \label{eq:cycle}
\end{equation}
For such a partition, $U_A=\exp(i\pi\sum_{i\in A}\mathcal J_i^z)$ reverses cross-sublattice exchanges and makes all off-diagonal entries nonpositive in the product-$m$ basis. Connectivity then supplies Perron--Frobenius uniqueness, and the LM spin comparison yields Eq.~\eqref{eq:lm}; the local term in Eq.~\eqref{eq:decompose} contributes only a scalar shift. Transporting a sublattice sign along a spanning tree and testing the remaining bonds implements the cycle condition~\cite{Harary1953}. The active graph contains sites with $k_i>0$, while disconnected components have additive spectra and couple by angular-momentum addition, potentially producing equal edges at different $K$. The bond-jump construction is explicit: with $b_{ij}=e_i+\operatorname{sgn}(J_{ij})e_j$, where $e_i$ is a site unit vector, its Gram matrix is $C=\sum_{i<j}|J_{ij}|b_{ij}b_{ij}^{\mathsf T}+\chi\bm1\bm1^{\mathsf T}+\operatorname{diag}(\kappa_i)$, a sum of positive semidefinite contributions.

For spin-$1/2$ sites, the bounds~\eqref{eq:globalcovbounds} also follow directly from this representation. The matrix inequality $C\succeq c_-\Id$ gives $-\LL_C\succeq(c_-/2)\sum_i\JJ_i^2$, hence a rate at least $qc_-$ in a module of $q$ active sites. On any set $I$ of $K$ sites, the highest-weight string $\prod_{i\in I}S_i^+$ is an exact spin-$K$ eigenoperator with rate $u(I)=[\operatorname{tr}C_I+\bm1^{\mathsf T}C_I\bm1]/2\leq Kc_+$, where $C_I$ is the principal Gram submatrix. Minimizing over these strings proves the global upper bound.

For the bond-noise chain with $\gamma>0$ and $L\geq3$, the stationary state follows from the Hermitian-jump identity
\begin{equation}
	\langle X,-\LL_CX\rangle_{\HS}
	=\frac12\sum_\mu\|[L_\mu,X]\|_{\HS}^2,
	\label{eq:stationarykernel}
\end{equation}
which identifies the kernel as the joint commutant of the jumps. Writing $J_i^a=S_i^a+S_{i+1}^a$, overlapping bonds give
\begin{equation}
	[J_i^a,J_{i+1}^b]=i\epsilon_{abc}S_{i+1}^c,
	\label{eq:bondcommutator}
\end{equation}
thereby generating every interior local spin, while subtraction from the bond sums recovers the endpoints. The resulting local-spin algebra has only scalar commutants, so $\rho_{\rm ss}=\Id/2^L$ is unique. Since the periodic chain contains this sufficient open-chain set of jumps and additional next-neighbor Hermitian jumps can only reduce its joint commutant, uniqueness holds throughout the frustrated family with nonzero nearest-neighbor noise.

\section{Symmetry-resolved calculation}
\label{app:numerics}

The effective spin-$1$ chain is represented in the product basis $m_i\in\{-1,0,1\}$ at fixed magnetization $M=\sum_i m_i$, with each exchange written as $\mathcal J_i^z\mathcal J_j^z+(\mathcal J_i^+\mathcal J_j^-+\mathcal J_i^-\mathcal J_j^+)/2$. The total raising operator maps this basis to the neighboring $M+1$ block, and its kernel in $M=K$ is exactly the highest-weight space of total operator spin $K$, which the sparse edge calculation selects by minimizing
\begin{equation}
	H_{\rm pen}^{(K)}=H+\mu\mathcal J_-\mathcal J_+
	\label{eq:penalty}
\end{equation}
in that magnetization block. A state of total operator spin $K'>K$ acquires the penalty $\mu[K'(K'+1)-K(K+1)]$, whereas the target highest-weight space is unchanged, so sufficiently large $\mu$ separates the desired edge from the higher-spin sectors. The reported energies use the unpenalized Hamiltonian, with $H\psi=E\psi$ and $\mathcal J_+\psi=0$ jointly testing the eigenstate and its spin assignment.

For normalized edge eigenoperators $R_{K,M}$ in Hilbert--Schmidt space, the boundary profile and periodic string correlation in Fig.~\ref{fig:haldane} are
\begin{equation}
 \begin{aligned}
 m_i&=\langle R_{1,1},\mathcal J_i^zR_{1,1}\rangle_{\HS},\\
 \mathcal O^z_{\rm str}(r)&=-\frac1L\sum_i
 \left\langle\mathcal J_i^z
 e^{i\pi\sum_{j=i+1}^{i+r-1}\mathcal J_j^z}
 \mathcal J_{i+r}^z\right\rangle_{R_{0,0}},
 \end{aligned}
 \label{eq:diagnostics}
\end{equation}
with periodic site labels taken modulo $L$. The nearest-neighbor calculation uses even open lengths up to $L=14$ and periodic lengths up to $L=12$, also evaluating the half-chain Schmidt spectrum in the periodic singlet, whose two virtual boundaries give the characteristic low-lying four-state structure~\cite{Pollmann2010}. We fit the open-chain splitting exponentially and compare the periodic gap and string order with established spin-$1$ benchmarks~\cite{WhiteHuse1993}; for the frustrated chain, roots of $E(K=1)-E(K=0)$ locate boundary crossings, while periodic diagnostics at $L=8,10,12$ track the bulk across the ordering-scan interval.

\section{Finite-size stability of the hierarchy}
\label{app:stability}

\subsection{Hermitian perturbations}

Consider a Hermitian decay representative $H(p)=H_0+pV$, with $p\geq0$ and $V$ preserving the chosen module and total spin, and let $v_K^-$ and $v_K^+$ denote the minimum and maximum eigenvalues of $V$ in exact sector $K$. The variational principle gives~\cite{Kato}
\begin{equation}
	\Gamma_K(0)+pv_K^-\leq\Gamma_K(p)
	\leq\Gamma_K(0)+pv_K^+.
	\label{eq:weyl}
\end{equation}
Subtracting the upper bound for $K$ from the lower bound for $K+1$ yields
\begin{equation}
	\Delta_K(p)\geq\Delta_K(0)-p a_K,
	\qquad a_K=v_K^+-v_{K+1}^-.
	\label{eq:stability}
\end{equation}
For a positive initial margin, Eq.~\eqref{eq:stability} guarantees persistence of the ordering for $p<\Delta_K(0)/a_K$ when $a_K>0$, and for all $p\geq0$ when $a_K\leq0$. Because the criterion uses only the initial spacing and the extremal perturbation energies in the two sectors, it remains available after the Marshall sign condition is lost. In the open frustrated chain, $V=H_2$ is the dimensionless perturbation, since the scalar shift $2\alpha(L-2)$ cancels from $\Delta_{01}/\gamma$. The exponentially small unperturbed boundary spacing and the extensive perturbation then narrow the guaranteed interval as the chain grows, while the actual crossing positions follow the effective interaction between the two boundary spins.

\subsection{Coherent perturbations}
\label{app:coherent}

A Hermitian SU(2)-scalar Hamiltonian preserves total operator spin while possibly coupling different local-rank patterns, so we work on the full exact-$K$ space and write $-\LL_\varepsilon=D_K+i\varepsilon A_K$, with $D_K=-\LL_C|_K$ Hermitian and $A_K=[H,\cdot]|_K=A_K^\dagger$. Let $P_K$ project onto the entire lowest eigenspace of $D_K$, with eigenvalue $d_K$, and set $Q_K=\Id-P_K$; when $Q_K\neq0$, denote the gap above this space by $g_K>0$ and the coupling between the two spaces by $b_K=|\varepsilon|\|Q_KA_KP_K\|$. To prove Eq.~\eqref{eq:coherentbound} for $b_K<g_K/2$, suppress the sector label and put $T=D+i\varepsilon A$. Any right eigenvector $v$ with eigenvalue $\mu$ gives the lower bound $\operatorname{Re}\mu=\langle v,Dv\rangle/\|v\|^2\geq d$, while an upper bound follows from the diagonal blocks $B_P=dP+i\varepsilon PAP$ and $B_Q=QDQ+i\varepsilon QAQ$. For $x=\operatorname{Re}\mu-d\in(0,g)$, their resolvents obey
\begin{equation}
	\|(B_P-\mu)^{-1}\|\leq\frac1x,\qquad
	\|(B_Q-\mu)^{-1}\|\leq\frac1{g-x}.
	\label{eq:coherentresolvents}
\end{equation}
The first inequality follows from normality of $B_P$ and the second from $\operatorname{Re}\langle q,(B_Q-\mu)q\rangle\geq(g-x)\|q\|^2$, so Schur complementation makes $T-\mu$ invertible whenever $b^2<x(g-x)$. Consequently the open vertical strip $d+r_-<\operatorname{Re}\mu<d+r_+$ contains no eigenvalues, where $r_\pm=(g\pm\sqrt{g^2-4b^2})/2$.

To count the eigenvalues on either side, multiply only the off-diagonal blocks $i\varepsilon PAQ$ and $i\varepsilon QAP$ by $t\in[0,1]$. The line $\operatorname{Re}\mu=d+g/2$ remains in the resolvent throughout this deformation because $tb<g/2$, and at $t=0$ precisely $\operatorname{rank}P$ eigenvalues lie to its left with real part $d$, while the other block has real parts at least $d+g$. Continuity preserves this count, including algebraic multiplicities, so the empty strip and the lower bound together confine the entire lowest cluster to $d\leq\operatorname{Re}\mu\leq d+r_-$. Rationalizing $r_-$ gives $r_-=2b^2/(g+\sqrt{g^2-4b^2})\leq2b^2/g$, proving the quadratic estimate for any degeneracy of $P$; if $Q=0$, all rates are exactly $d$. Applying this upper bound in sector $K$ and the unconditional lower bound in sector $K+1$ yields $\Gamma_\varepsilon(K+1)-\Gamma_\varepsilon(K)\geq d_{K+1}-d_K-r_K$, so an initial positive margin survives whenever it exceeds $r_K$. The estimate controls only the coupling out of the slow eigenspace, allowing arbitrarily strong coherent dynamics within either diagonal block, and the same proof holds in a refined module whenever both the noise and the coherent dynamics preserve it.

\section{Bond Berry phase}
\label{app:berry}

In the periodic representative $H_1+\alpha H_2$, twist one nearest-neighbor bond $(i,j)$ by replacing its transverse exchange with
\begin{equation}
	\frac12\left(e^{i\theta}\mathcal J_i^+\mathcal J_j^-
	+e^{-i\theta}\mathcal J_i^-\mathcal J_j^+\right).
	\label{eq:twist}
\end{equation}
With its longitudinal term and all other bonds unchanged, the loop preserves $M$ but generally not total $K$, so we follow the normalized lowest state $|\psi(\theta)\rangle$ in $M=0$, continuing the singlet at $\theta=0$. For a nondegenerate closed loop, the discretized Wilson product and Berry phase are
\begin{align}
	W&=\prod_{j=0}^{N_\theta-1}
	\frac{\langle\psi(\theta_j)|\psi(\theta_{j+1})\rangle}
	{|\langle\psi(\theta_j)|\psi(\theta_{j+1})\rangle|},\notag\\
	\gamma_B&=-\arg W\pmod{2\pi},\qquad
	\theta_j=\frac{2\pi j}{N_\theta},
	\label{eq:berry}
\end{align}
with $\theta_{N_\theta}=\theta_0$, so the normalized overlap product removes arbitrary eigenvector phases. Integer-spin time reversal leaves the twisted Hamiltonian invariant at each $\theta$ and satisfies $\mathcal T^2=1$, making the nondegenerate state a real line bundle whose holonomy is $+1$ or $-1$ and hence quantizing $\gamma_B$ to $0$ or $\pi$~\cite{Hatsugai2006Berry}. We use $L=6,8,10$ and $N_\theta=32$, with Fig.~\ref{fig:frustration}(b) showing the two larger sizes; the two lowest $M=0$ eigenvalues give the minimum sampled sector gap, while successive overlaps monitor continuation on the angular mesh. Comparing lengths then distinguishes the finite-size drift of the bond index from the open-chain boundary crossings.

\section{Projected probes and absolute signals}
\label{app:probe}

\subsection{A local two-spin readout}
\label{app:dimerreadout}

Write $P_{ab}=\sigma_1^a\sigma_2^b$ and choose
\begin{align}
	B_0&=(P_{xx}+P_{yy}+P_{zz})/(2\sqrt3),\notag\\
	B_1&=(P_{xy}-P_{yx})/(2\sqrt2),\notag\\
	B_2&=(P_{xy}+P_{yx})/(2\sqrt2).
	\label{eq:dimerprobes}
\end{align}
These Hermitian traceless bilinears obey $\Tr(B_KB_{K'})=\delta_{KK'}$ and $\JJ^2B_K=K(K+1)B_K$, and their local ranks are both one, so Eq.~\eqref{eq:dimer} gives $\LL_C(B_K)=-\Gamma_{(1,1)}(K)B_K$. To establish separability of the preparations in Eq.~\eqref{eq:dimerreadout}, write any state $\rho=(\Id+\sum_{ab}t_{ab}P_{ab})/4$ with $\sum_{ab}|t_{ab}|\leq1$ as a convex mixture of $\Id/4$ and $\Omega_{ab}^{\pm}=(\Id\pm P_{ab})/4$, with weights $1-\sum_{ab}|t_{ab}|$ and $|t_{ab}|$ and signs selected by $t_{ab}$. Each $\Omega_{ab}^{\pm}$ is an equal mixture of two product states whose local eigenvalues along axes $a,b$ are aligned or antialigned. The sums of absolute coefficients, $2\sqrt3\epsilon_{\rm p}$ for $\rho_0$ and $2\sqrt2\epsilon_{\rm p}$ for $\rho_{1,2}$, establish the common preparation bound, while unitality gives
\begin{equation}
	e^{\LL_Ct}\rho_K(0)=\Id/4+
	\epsilon_{\rm p}e^{-\Gamma_{(1,1)}(K)t}B_K,
\end{equation}
which proves Eq.~\eqref{eq:dimerreadout}. Since the deviation remains proportional to $B_K$, both the normalized Hilbert--Schmidt and trace distances follow the same exponential. The symbols in Fig.~\ref{fig:dimer}(b) independently follow by exponentiating the full $16\times16$ Liouvillian built from the six physical jumps $L_{\pm,a}$, with trace, positivity, and sector checks on the resulting density matrices.

\subsection{Many-spin projected probes}

For unit local vectors $\bm n_i$, define a unit-norm full-support string
\begin{equation}
	\widehat P=2^{-L/2}\bigotimes_i(\bm n_i\cdot\bm\sigma_i).
	\label{eq:pauli}
\end{equation}
All local ranks are one, and the chosen axes repeat as $\bm x$, $(\bm x+\bm y)/\sqrt2$, $(\bm y+\bm z)/\sqrt2$, and $(\bm z+\bm x)/\sqrt2$. The exact highest-weight projector is the SU(2) group matrix unit
\begin{equation}
	\mathcal P_{KK}^{(K)}(X)=(2K+1)\int dg\,
	D_{KK}^{(K)}(g)^*U_gXU_g^\dagger,
	\label{eq:projector}
\end{equation}
with normalized Haar measure. Schur orthogonality projects onto $M=K$ in every copy of the spin-$K$ representation. Set
\begin{equation}
	q_K=\|\mathcal P_{KK}^{(K)}\widehat P\|_{\HS}^2,
	\qquad O_K=\mathcal P_{KK}^{(K)}\widehat P/\sqrt{q_K}.
	\label{eq:probe}
\end{equation}
The $M=K$ probe $O_K$ is used in Fig.~\ref{fig:dynamics}; in the fixed-magnetization basis, it is obtained by selecting the $M=K$ component of $\widehat P$ and projecting onto the kernel of the total raising matrix $J_+$ through
\begin{equation}
	P_{\ker J_+}=\Id-J_+^\dagger(J_+J_+^\dagger)^+J_+,
	\label{eq:kernel}
\end{equation}
where $+$ on the parenthesis denotes the Moore--Penrose inverse. This realizes the group projection~\eqref{eq:projector} and gives the edge weight $w_{K,0}=|\langle R_{K,K},O_K\rangle_{\HS}|^2$ for a nondegenerate highest-weight edge, with squared overlaps summed over the eigenspace when the edge is degenerate.

The product initial state
\begin{equation}
	\rho(0)=\bigotimes_i\frac{\Id+p\bm n_i\cdot\bm\sigma_i}{2},
	\qquad 0<p\leq1,
	\label{eq:productstate}
\end{equation}
contains the full-support term $p^L2^{-L/2}\widehat P$, while every other term has a different conserved rank pattern and is orthogonal to $O_K$, so covariance and Eq.~\eqref{eq:projector} give the measured amplitude
\begin{align}
	A_K(t)&=\Tr[O_K^\dagger e^{\LL_Ct}\rho(0)]\notag\\
	&=p^L2^{-L/2}\sqrt{q_K}\,C_K(t).
	\label{eq:amplitude}
\end{align}
For $K>0$, $O_K$ is generally non-Hermitian, so its complex expectation is reconstructed as $A_K=\langle X_K\rangle-i\langle Y_K\rangle$ from the Hermitian quadratures $X_K=(O_K+O_K^\dagger)/2$ and $Y_K=(O_K-O_K^\dagger)/(2i)$, whose spin projection on the full-support string can be implemented by global-rotation averages.

The approach to single-rate relaxation follows by writing $C_K(t)=e^{-\Gamma_Kt}[w_K+\sum_{n>0}w_{K,n}e^{-\delta_{K,n}t}]$, where $\delta_{K,n}=\Gamma_{K,n}-\Gamma_K>0$ on the supported excited modes, and using their positive weights to obtain Eq.~\eqref{eq:visibilitybound}. Its relative correction is at most $\varepsilon>0$ once $t\geq\delta_K^{-1}[\ln((1-w_K)/(\varepsilon w_K))]_+$, where $\delta_K$ is the smallest supported spacing and $0<w_K<1$; Eq.~\eqref{eq:amplitude} then connects this criterion to the prepared signal. Time propagation uses a Krylov representation of the Hermitian decay operator with the sector edge subtracted to control the large exponential scale, after which restoring $e^{-\Gamma_Kt}$ gives the normalized correlation and its physical logarithmic decay rate.

\section{Strong symmetry, exchange noise, and Hubbard loss}
\label{app:strong}

Under column vectorization the generator in Eq.~\eqref{eq:lindblad} has the matrix representation
\begin{align}
	\mathbb L={}&-i(\Id\otimes H-H^{\mathsf T}\otimes\Id)\notag\\
	&+\sum_\mu\left[L_\mu^*\otimes L_\mu
	-\frac12\Id\otimes L_\mu^\dagger L_\mu
	-\frac12(L_\mu^\dagger L_\mu)^{\mathsf T}\otimes\Id\right].
	\label{eq:vectorized}
\end{align}

\subsection{Physical-spin resolution of operator space}

When $[H,S_{\rm tot}^a]=[L_\mu,S_{\rm tot}^a]=0$, decompose physical states as $\mathcal H=\bigoplus_S(M_S\otimes V_S)$. Schur's lemma gives
\begin{align}
	\mathcal B(\mathcal H)&=\bigoplus_{S_L,S_R}
	(M_{S_L}\otimes M_{S_R}^*)\otimes(V_{S_L}\otimes V_{S_R}^*),\notag\\
	\LL|_{S_L,S_R}&=B_{S_L,S_R}\otimes\Id,\qquad
	V_{S_L}\otimes V_{S_R}^*\simeq
	\bigoplus_{K=|S_L-S_R|}^{S_L+S_R}V_K.
	\label{eq:strongblocks}
\end{align}
Within a fixed $(S_L,S_R)$ block, each multiplicity-space eigenvalue is repeated throughout the indicated range of $K$, giving equal sector edges there. A particle-number grading can resolve the blocks further. After the stationary and oscillatory nondecaying modes are removed, the smallest strictly positive rate $\Gamma^{(+)}$ describes relaxation toward the associated stationary or decoherence-free subspace~\cite{BucaProsen2012,AlbertJiang2014}.

\subsection{Exchange noise and SU(4) branching}
\label{app:su4}

For the spin-$1/2$ exchange jump in Sec.~\ref{sec:young}, $P_i^2=\Id$ implies
\begin{equation}
	[h_i,[h_i,X]]=\tfrac12(X-P_iXP_i).
	\label{eq:swapdouble}
\end{equation}
Vectorization turns $P_iXP_i$ into the simultaneous swap of the two Liouville legs, $\Pi_i^{(4)}=P_i^LP_i^R$, proving Eq.~\eqref{eq:su4} and embedding the physical subgroup as $[\mathrm{SU}(2)_L\times\mathrm{SU}(2)_R]/\mathbb Z_2\subset\mathrm{SU}(4)$. For uniform $\kappa$, the resulting permutation chain is Bethe integrable~\cite{Sutherland1975}, with a one-color defect $|j\rangle$ in a uniform background obeying
\begin{equation}
	-\LL_{\rm ex}|j\rangle=\frac\kappa2
	(2|j\rangle-|j-1\rangle-|j+1\rangle).
	\label{eq:su4defect}
\end{equation}
Fourier transformation gives the exact periodic branch $\Gamma(q)=\kappa(1-\cos q)$, with $q=2\pi m/L$ and a smallest nonzero rate asymptotic to $2\pi^2\kappa/L^2$. A physical $(S_L,S_R)$ edge is obtained by branching the contributing SU(4) representations into the physical subgroup and minimizing over their multiplicity eigenvalues. This connects the Young-diagram organization of the exchange generator to its physical-spin relaxation channels.

\subsection{Complex-interaction Hubbard spectrum}
\label{app:hubbard}

For Eq.~\eqref{eq:hubbard}, the effective Hamiltonian between jumps is
\begin{equation}
	H_{\rm eff}=H_{\rm Hub}-\frac{i\kappa}{2}
	\sum_i n_{i\uparrow}n_{i\downarrow}.
	\label{eq:heffhubbard}
\end{equation}
Pair recycling maps $(N_L,N_R)$ to $(N_L-2,N_R-2)$, making the finite Liouvillian block triangular. Its diagonal blocks $-i(H_{\rm eff}X-XH_{\rm eff}^\dagger)$ determine the full eigenvalue multiset through Eq.~\eqref{eq:hubbardwidth}~\cite{Nakagawa2021}, while recycling fixes how the corresponding eigenoperators extend across particle-number sectors and can modify their Jordan structure.

For the periodic chain, put $u=(U-i\kappa/2)/(4t_{\rm h})$ and $\Theta(x)=2\arctan x$. The analytically continued Lieb--Wu equations may be written
\begin{align}
	k_jL&=2\pi I_j-\sum_{\alpha=1}^{M}
	\Theta\!\left(\frac{\sin k_j-\lambda_\alpha}{u}\right),\notag\\
	-\sum_{j=1}^N\Theta\!\left(\frac{\sin k_j-\lambda_\alpha}{u}\right)
	&=2\pi J_\alpha+\sum_{\beta=1}^{M}
	\Theta\!\left(\frac{\lambda_\alpha-\lambda_\beta}{2u}\right),
	\label{eq:bethe}
\end{align}
with energy $E^{\rm eff}=-2t_{\rm h}\sum_j\cos k_j$. The Bethe roots and logarithmic branches are continued from admissible quantum numbers~\cite{LiebWu1968,Nakagawa2021}, with $M=N/2-S$ in a highest-weight spin sector. The spin-resolved loss width is consequently
\begin{equation}
	\kappa_{\min}(N,S)=\min_{a\in(N,S)}[-2\operatorname{Im}E_a^{\rm eff}],
	\label{eq:hubbardminwidth}
\end{equation}
where the minimization includes the relevant Bethe descendants and limiting solutions. Combining these widths on the two Liouville legs through Eq.~\eqref{eq:hubbardwidth} gives the corresponding decay edges.

For $0<N/L\leq1$, the maximal-spin sector $S=N/2$ has no onsite singlet pairs and is dark, while the large-$L$, one-spin-rapidity branch at fixed density and consecutive charge quantum numbers gives the small-momentum dark--decaying coherence rate~\cite{Nakagawa2021}
\begin{multline}
	\Gamma_{\rm sw}(P)\simeq
	\frac{2t_{\rm h}^2\kappa}{\pi[U^2+(\kappa/2)^2]}
	\left(Q_0-\tfrac12\sin2Q_0\right)\\
	{}\times\left[1-\cos\!\left(\frac{\pi P}{Q_0}\right)\right]
	\simeq D_sP^2,\qquad Q_0=\pi N/L.
	\label{eq:spinwave}
\end{multline}
Here $D_s$ is linear in $\kappa$ at weak loss for fixed $U\neq0$ and crosses to quantum-Zeno scaling $D_s\propto1/\kappa$ at strong loss, while the corresponding population decays at twice the coherence rate by the two-leg sum in Eq.~\eqref{eq:hubbardwidth}. The two-component fermionic Yang--Gaudin gas with contact singlet loss supplies a continuum counterpart, whose nested Bethe ansatz~\cite{Yang1967,Gaudin1967} admits a complex-coupling extension and allows spin-resolved lifetimes to be compared through complex-energy widths; recent calculations identify loss-dependent changes in spin stability among selected few-body Bethe configurations~\cite{KatsutaUchino2026}.

\subsection{Nonnormal spectral stability}

The Hermitian representatives of the noise chain admit the variational treatment in Appendix~\ref{app:numerics}, while Arnoldi iteration resolves the complex spectrum of a general Lindblad block~\cite{Arnoldi1951}, whose first-positive edge is extracted after separating the peripheral subspace. In this setting, eigenvector conditioning enters spectral stability: for a diagonalizable block $\LL_r=R\Lambda R^{-1}$, every eigenvalue of $\LL_r+\delta\LL_r$ lies within distance $\|R\|\|R^{-1}\|\|\delta\LL_r\|$ of the unperturbed spectrum~\cite{TrefethenEmbree}, complementing the direct Hermitian bound of Appendix~\ref{app:stability} with the conditioning factor $\|R\|\|R^{-1}\|$.

\bibliography{references}
\end{document}